# Thickness driven topological quantum phase transition and experimental evidence of Dirac semimetal state in Cr-doped Bi$_2$Se$_3$

Turgut Yilmaz[1,*]

[1]Department of Physics, Science and Literature Faculty, Uludag University, Bursa 16059, Turkey

[*]E-mail: turgutyilmaz@uludag.edu.tr

Date: 10/26/2017

**In this report, we demonstrate that the energy gap in Cr-doped Bi$_2$Se$_3$ closes and reopens in oscillatory fashion with increasing sample thickness indicating the topological quantum phase transition driven by the quantum finite size effect. The pattern of the oscillation provides an evidence that Cr-doped Bi$_2$Se$_3$ can have a Dirac semimetal state depending on the sample thickness and Cr-content. This resolves the puzzle of non-magnetic gap opening at the Dirac point of topological insulators. An energy gap can be opened due to the quantum finite size confinement of the bands in Dirac semimetal state. In addition, our room temperature ARPES measurements revealed a topological quantum phase transition between x = 0.06 and x = 0.08 in 7 quintuple layers Bi$_{2-x}$Cr$_x$Se$_3$ without ferromagnetism.**

In recent years, magnetic impurity-doped topological insulators (TIs) have gained tremendous attention by the scientific community due to being a host material for experimental observation of the quantum anomalous Hall effect (QAHE) [1-2]. In the quantum anomalous Hall state, longitudinal Hall resistance ($\rho_{xx}$) drops to zero, while transverse hall resistance ($\rho_{xy}$) quantized in the unit h/(ne$^2$) where n is an integer number, h is the Planck's constant and e is the electron charge. This is the consequence of the one-way charge carrying dissipationless edge states located at the sample edges. Therefore, QAHE has a potential for future electronic devices with low-power consumption and high speed. However, extremely low observation temperature of the QAHE hinders the device developments. In the recent experiments, QAHE has been observed in V- and Cr-doped (Bi, Sb)$_2$Te$_3$ in the tens of K range, while the ferromagnetic transition temperature is ~30 K [2-4].

One of the intriguing observation is that the QAHE can be observed in the narrow range of the sample thickness. A recent work showed that the quantum anomalous hall state demonstrates thickness dependence in Cr-doped (Bi,Sb)$_2$Te$_3$ that it disappears above and the below 5 Quintuple layers (QL) Cr$_{0.15}$(Bi$_{0.1}$Sb$_{0.9}$)$_{1.85}$Te$_3$ [5]. This is attributed competition between the exchange field and the hybridization induced energy gap for below 5 QL and the larger thickness than the dephasing length for above 5 QL. However, another study showed that the QAHE can be also observed at 10 QL regardless of the dephasing length [6].

The sample thickness not only has the impact on the Hall response of the samples but also modulates the electronic structure. The energy gap in a thin TI oscillates with an exponential decays as a function of thickness and closes after few QL [7-9]. The oscillatory pattern of the energy gap has been also theoretically purposed for another topological material known as Dirac semimetals [10-11]. Dirac semimetals (DSMs) have 3D Dirac cone formed by the non-trivial bulk bands [12]. In DSMs, the energy gap oscillates with increasing sample thickness by passing an energy gap minimum, while the energy gap maximum decays. Therefore, both topological materials show a distinct evolution of the electronic structure as a function of the sample thickness. Up to our knowledge, these results have not been experimentally observed in the band structure experiments.

The absence of the QAHE below and above a certain thickness can be understood from the evolution of the electronic structure as a function of the sample thickness. Because across the topological quantum phase transition (TQPT) between the topologically distinct materials an energy gap closing is expected. This can be resolved by angle resolved photoemission spectroscopy (ARPES) [13]. Herein, we investigate the electronic structure of Cr-doped $Bi_2Se_3$ as a porotype sample by varying the thickness and the Cr-content. Remarkably, we observed that the energy gap in $Bi_{1.98}Cr_{0.02}Se_3$ closes at the critical thicknesses and reopens in oscillatory fashion with increasing sample thickness. This indicates that Cr-doped samples at different thicknesses are topologically distinct materials, which explains the dependence of the QAHE on the sample thickness observed in Reference 5. Furthermore, such pattern of the energy gap oscillation suggests that Cr doped $Bi_2Se_3$ can be a DSM depending on the sample thickness and Cr-content. This also reveals the origin of the non-magnetic gap opening in Cr-doped $Bi_2Se_3$, which still remains as a puzzle in the literature. In the non-magnetic state, the energy gap opens due to finite size confinement of the energy bands along to the surface normal in DSM state. In addition, our room temperature ARPES measurements revealed a TQPT between x = 0.06 and x = 0.08 in 7 QL $Bi_{2-x}Cr_xSe_3$ without ferromagnetism. This observation is in contrast to a recent work that claimed to the presence of a magnetic-field-driven TQPT in $Bi_{2-x}Cr_xSe_3$ at x = 0.07 [14]. Hence, our experiments carry rich information and resolve the several unsolved-questions in TI research area.

High quality pristine and Cr-doped $Bi_2Se_3$ samples were grown on $Al_2O_3$ (0001) substrate by using molecular beam epitaxial method (MBE) and 5N purity sources. Growth and photoemission chambers were attached each other that photoemission experiment can be performed on a freshly grown sample without breaking the vacuum. Scienta SES100 electron analyzer with an energy resolution of 25 meV and angular resolution better than $0.2^o$ was used to record ARPES maps. He I$\alpha$ and Xe I$\alpha$ lines with photon energies of 21.2 eV and 8.437 eV, respectively were used the excite electrons for photoemission process. Vacuums are $1x10^{-9}$ torr in the MBE and $2x10^{-10}$ torr in the photoemission chamber.

To investigate the finite size effect on the electronic structure of Cr-doped $Bi_2Se_3$, we presented ARPES maps of $Bi_{1.98}Cr_{0.02}Se_3$ within the thickness range of 7 QL and 12 QL in Figure 1(a) along with the electronic structure of a pristine sample. Pristine $Bi_2Se_3$ exhibits well-resolved surface states with the Dirac point located at 310 meV binding energy and at Γ (000). Upon Cr incorporation into the bulk, an 80 meV energy gap was observed at the DP of the 7 QL film. The energy gap follows progressive drop with increasing sample thickness and remarkably vanishes at 12 QL in which the gapless linear surface states are formed reflecting the topological character of the material. Since there is no energy gap closing or reversing up to 12 QL, the TQPT is absent within the thickness range of 7QL and 12 QL.

In order to draw a bigger picture, we presented the electronic structure of $Bi_{1.98}Cr_{0.02}Se_3$ for further increased sample thicknesses in Figure 1(b). Increasing thickness from 12 QL to 20 QL leads to reopening a 40 meV energy gap at the DP, which is smaller than the energy gap measured at 7 QL. Growing the thicker film suppresses the energy gap, which drops to 20 meV at 24 QL film. When the sample thickness reaches to 27 QL, the energy gap vanishes again by forming the Dirac surface states as shown in Figure 1(b). Furthermore, a 20 meV energy gap opens again with further increasing the thickness to 30 QL. Hence, Our spectroscopic study demonstrates that the energy gap oscillates as a function of sample thickness. We noted that the Cr-content is carefully adjusted and fixed during the increasing sample thickness. Therefore, the oscillation of the energy gap is a consequence of the varied sample thicknesses. In addition to the oscillation of the energy gap with the sample thickness, another notable observation is that the gap at the thicker film is much smaller than the one measured at the thinner film indicating the decay of the energy gap size with increasing thickness. The gap is measured to be 80 meV at 7 QL, while it is 40 meV at 20 QL and 20 meV at 30 QL of the film. This reveals that increasing thickness leads to the smaller energy gap at the DP.

We also compared the electronic structure of the 12 QL and 27 QL films in which the gapless surface states are formed (Figure 2). First, the increasing thickness has hole doping effect on the electronic structure. The binding energy of the DP is 290 meV for 12 QL and 270 meV for 27 QL indicating that the DP shifts to the lower binding energy with increasing thickness. Furthermore, the surface states and bulk bands overlap stronger with increasing thickness as shown in Figure 2. High-intensity points are marked with blue and red solid lines in Figure 2(c) for 12 QL and 27 QL films, respectively. As seen in Figure 2(c), the high-intensity point of the 27 QL is not well resolved compared to the 12 QL film due to the stronger overlapping of the surface states and the bulk bands. This shows that bulk band gap decreases with increasing sample thickness.

We extended our study and analysis the increasing Cr-content impact on the electronic structure of 7 QL and 27 QL $Bi_2Se_3$. In Figure 3(a), ARPES maps of 27 QL are presented $Bi_{2-x}Cr_xSe_3$ for different Cr-content. The pristine sample has gapless Dirac like surface states with the DP located at 190 meV and at Γ (000). Introducing Cr-content of $x = 0.02$ into the bulk does not alter the electronic structure except for the shift of DP to the higher binding energy from 190 meV to 270 meV. Further increasing Cr-content leads to open an energy gap and the gap size increases with higher Cr-content. Since the amplitude of the gap size does not reverse with increasing Cr-content from $x = 0$ to $x = 0.06$ QTPT is absent within this Cr concentration for the 27 QL film. Furthermore, ARPES maps of 7 QL film are also presented in Figure 3(b) for increasing Cr concentration. The energy gap in 7 QL film exhibits a quite different and important pattern as a function of Cr-content. The 80 meV energy gap drops to 55 meV first for Cr content of $x = 0.04$ and does not show visible changes for increasing Cr content to $x = 0.06$. Then, the energy gap enlarges to 80 meV when the Cr-content reaches to $x = 0.08$ that indicates the TQPT is taking place between $x = 0.06$ and $x = 0.08$. This observation is remarkable consistent with a recent transport measurement experiment [14]. It was shown that the increasing Cr content in $Bi_2Se_3$ leads to a crossover between weak antilocalization to weak localization at $x = 0.07$ [14]. This refers to the TQPT from TI to a normal insulator (NI), which is claimed to be due to the magnetization [14]. However, our work showed that the transition is also taking place at the room temperature regardless of magnetism since the ferromagnetic transition temperature of Cr-doped samples is ~20 K [16].

In the next step, we investigate the evolution of the electronic structure of Cr-doped $Bi_2Se_3$ as a function of the sample thickness with different photon energy. In the photoemission experiment, the electron intensity of the surface and the bulk states exhibit different photon energy dependence. Bulk bands strongly disperse as a function of the photon energy, while the surface states show negligible dependence [17]. By using this fact, the surface states can be distinguished in ARPES by varying the incident photon energy by eliminating the bulk bands. Therefore, we carried out ARPES with Xe Iα radiation (hν = 8.437 eV). In figure 3, ARPES maps of $Bi_{1.98}Cr_{0.02}Se_3$ are given as a function of thickness. Due to the matrix element effect, the bulk bands almost disappears in the spectra leaving well-resolved surface states. Similar to the ARPES maps taken with 21.2 eV photon energy, the gap vanishes at 12 QL and reopens with increasing thickness indicating the TQPT.

The energy gap versus the sample thickness is summarized in Figure 5(a) that the gap oscillates with increasing sample thickness by passing gap minimums at zero (black squares) and decaying gap maximums (blue squares). The band gap closing at the critical points in which Z2 invariant number changes between 0 (NI) and 1 (TI) refers TQPT [18-22]. Hence, we conclude that 12 QL and 27 QL are the critical thicknesses between the topologically distinct materials and showed the thickness-driven TQPT in Cr-doped $Bi_2Se_3$.

Furthermore, the oscillation of the energy gap in our samples is distinct than TIs. Frist of all, the energy gap in TIs has oscillatory behavior up to only a few QLs and the oscillation of the energy gap is extremely small (<1meV) [7-9] compared to the pattern observed in our work. The oscillation in our sample can be easily resolved even up to 30 QL film. In addition, the oscillation in our samples accompanies with critical thicknesses in which the energy gap reaches the minimum and closes. However, the energy gap always exists in the oscillatory pattern of TIs [8]. Only the energy gap maximum marked with A, B, C in Figure 5(a) decays in our case. Moreover, in contrast to the persistence of the gapless surface states in TIs in and above 6 QL [15], the energy gap in our sample reopens even above 27 QL thickness. Hence, we conclude that the energy gap in Cr-doped $Bi_2Se_3$ shows distinct oscillatory pattern than TIs.

The similar oscillatory pattern to ours is theoretically predicted for $Cd_3As_2$ DSM [10-11]. The oscillation of the energy gap in DSM also possesses an energy gap minimum and the decaying gap maximum. Furthermore, the gap can be opened at the DP of DSM due to the confinement of the energy bands even at the much thicker film than 6 QL [10]. Therefore, the oscillatory pattern of the energy gap suggests that Cr-doped $Bi_2Se_3$ can be a DSM. This clarifies a puzzling question in the literature that Cr- and Mn-doping open an energy gap at the DP of $Bi_2Se_3$ without requiring a magnetization [23-24]. This is in contrast to expectations that the topological surface states are protected against nonmagnetic perturbations. Our results show that Cr-doped $Bi_2Se_3$ has an energy gap at the DP due to being a DSM since the finite size effect can open an energy gap at the DP of a DSM even at the thicker films [10]. In addition, this also explains the broader energy bands in Cr-doped $Bi_2Se_3$, which are the characteristics of DSMs due to possessing of 3D bulk bands.

Moreover, Cr-doping does not open an energy gap and broaden the energy bands in all TIs. Such as, (Bi, Sb)$_2$Te$_3$ exhibits gapless sharp surface states even against heavy Cr-bulk doping [27]. The difference between $Bi_2Se_3$ and (Bi, Sb)$_2$Te$_3$ is the strength of the spin-orbit-coupling (SOC), which is the key factor behind the band inversion in TIs. (Bi, Sb)$_2$Te$_3$ is expected to have a larger SOC than $Bi_2Se_3$ since Se has much smaller SOC than Te and Bi. Therefore, Cr-substitution with Bi in $Bi_2Se_3$ will have a more dramatic impact on the electronic structure. In addition, a TI material can be turned into a DSM by closing the bulk band gap through the weakening of the SOC strength. Therefore, we purpose that Cr-doping drives $Bi_2Se_3$ into the DSMs by weakening the SOC strength. As a result, energy bands will be 3D as observed broader energy bands in our work and an energy gap opens at the DP due to the finite size effect in Cr-doped $Bi_2Se_3$.This is in agreement with a recent theoretical calculation that Cr-doping turns $Bi_2Se_3$ into an NI due to the weakened SOC strength [28]. In Cr-doped (Bi, Sb)$_2$Te$_3$, however, the presence of the Te in the compound ensure the TI phase due to its large SOC strength. Hence, our approach also explains the non-magnetic gap opening as well as the broader energy bands in the impurity-doped TIs. This also elucidates the TQPT with increasing Cr-content at the 7 QL $Bi_{2-x}Cr_xSe_3$ that TQPT takes places due to the completion between the finite sizes induced gap and the decaying of the gap due to the weakened SOC.

In addition, our spectroscopic study suggests that the gapless DSM can be also obtained from Cr-doped $Bi_2Se_3$ with a 3D bulk Dirac cone by further increasing the sample thickness and consequently suppressing the bulk band confinement induced gap. Experiments can be further carried to induce a ferromagnetism in order to break TRS. This will separate the overlapped DP in momentum states and 4 non-degenerate bulk Dirac cones can be obtained [25]. The material with such electronic structure is known as Weyl semimetal (WSM) [12].

In conclusion, we experimentally provided a strong evidence that Cr-doped $Bi_2Se_3$ is a DSM. Increasing thickness of the sample leads a very similar oscillatory pattern in the energy gap as purposed for DSMs. Furthermore, our observation resolves the non-magnetic gap opening and broader energy bands in the transition metal-doped TIs. Due to the confinement of the energy bands, a gap emerges at the DP and the broader bands are the results of the characteristic bulk bands of DSM state. Finally, we purposed a strategy to obtain a DSM with a gapless 3D Dirac cone and a WSM from a TI by growing much thicker film and inducing a net out-of-plane magnetic moment.

We noted that there could be more critical thicknesses where the gap closes or reverses in its magnitude. More detailed theoretical and experimental works can be done for further analysis to determine of Z2 invariant number for different thicknesses and Cr-content since it is not possible to resolve the topology of the bulk bands with ARPES. Nevertheless, our observations are important by itself that they carry rich information regarding the topological phase transition, the non-magnetic energy gap opening at DP of TIs and the thickness dependence of the QAHE. Therefore, our results can guide future studies in order to obtain QAHE at higher temperatures with even higher Chern numbers [26].

**AUTHOR INFORMATION**
Corresponding Author
*turgutyilmaz@uludag.edu.tr
Present Addresses:
§ Department of Physics, Faculty of Arts and Science, Uludag University, Gorukle Campus, Bursa, 16059, Turkey.

**Notes**
The author would like to thank Prof. Boris Sinkovic for useful discussions and experimental supports.
**ACKNOWLEDGMENT**
No potential conflict of interest was reported by the author and this work was funded by the University of Connecticut un-der the UCONN-REP (Grant No. 4626510).

**Figures:**

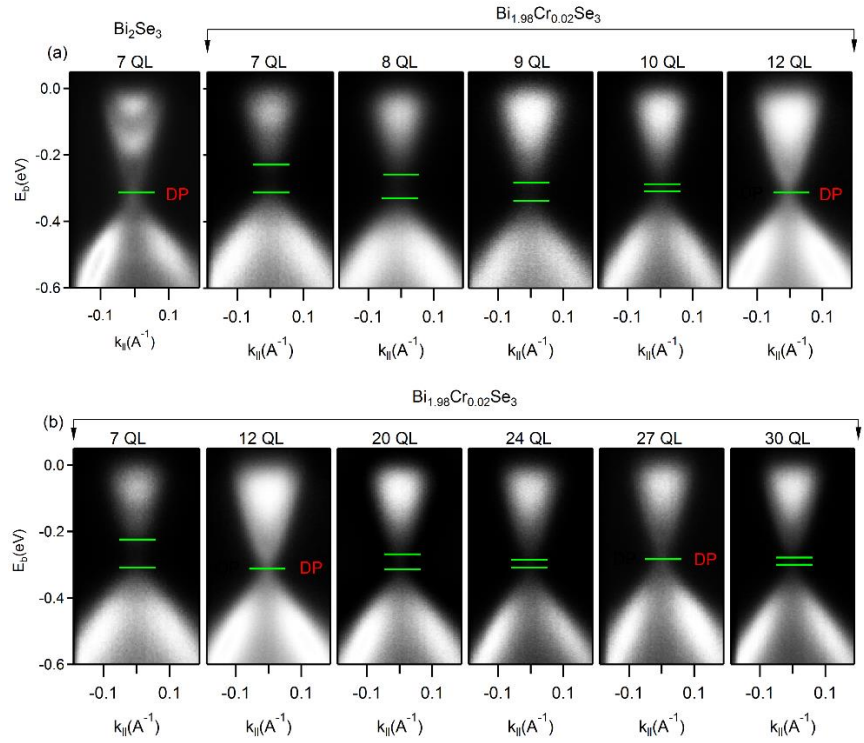

**Figure 1. (a) – (b)** ARPES map of a 7 QL $Bi_2Se_3$ and thickness dependent electronic structure of $Bi_{1.98}Cr_{0.02}Se_3$. Solid green lines represent the energy gap. Sample thicknesses are given on the top of the each map. Spectra were obtained at room temperature along Γ – M direction with He Iα photons.

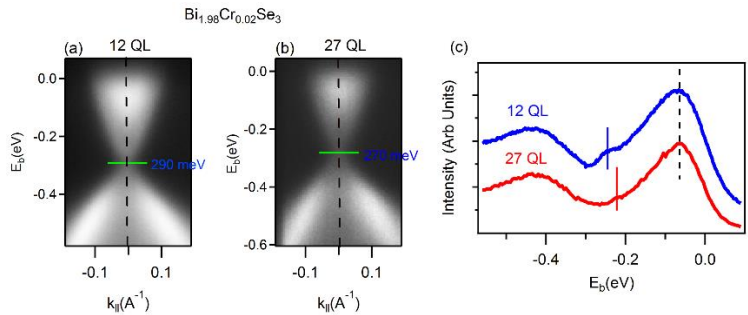

**Figure 2. (a) – (b)** ARPES maps of 12 QL and 27 QL $Bi_{1.98}Cr_{0.02}Se_3$, respectively. (c) Energy dispersive curves along the dashed black lines in (a) and (b). Solid blue and red lines represent the high-intensity points of the surface states.

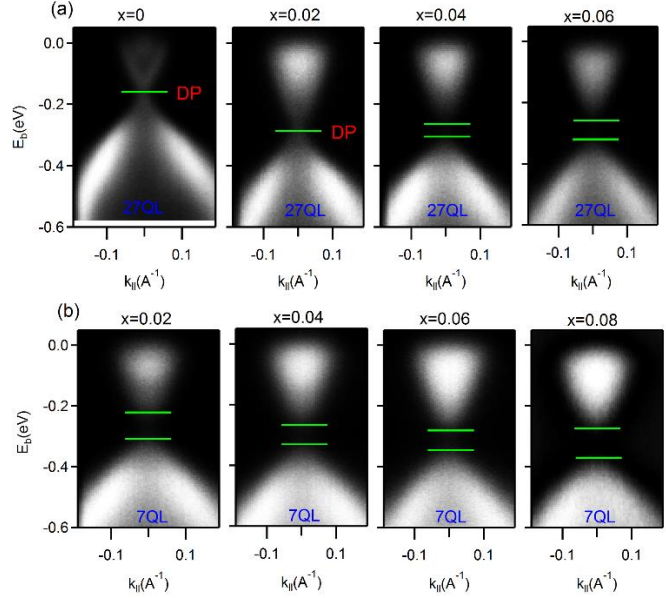

**Figure 3.** **(a) – (b)** ARPES maps of 27 QL and 7 QL $Bi_{2-x}Cr_xSe_3$ as a function of Cr-content. Solid green lines represent the energy gap. Cr-contents are written on the top of the each spectrum.

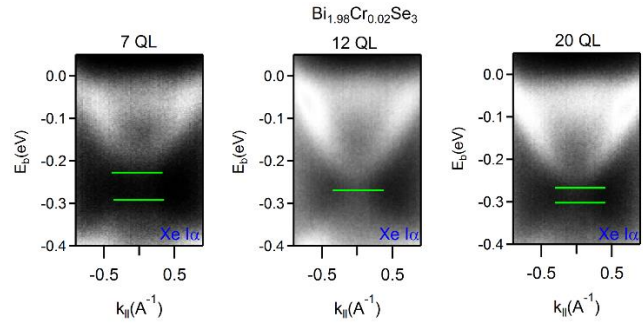

Figure 4. ARPES maps of $Bi_{1.98}Cr_{0.02}Se_3$ are given as a function of thickness obtained with Xe Iα radiation (hν = 8.437 eV).

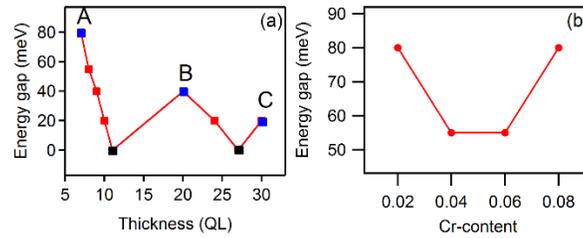

**Figure 5.** (a) Energy gap as a function of the thickness. Blue and black squares represent the gap maximums and minimums, respectively. (b) Cr-content dependence of the energy gap in 7 QL film.

8